\documentclass[10pt]{iopart}

\usepackage{graphicx}
\usepackage[dvips]{color}

\begin{document}

\title{Transition from Resonances to Bound States in Nonlinear Systems:
Application to Bose-Einstein condensates}
\author{Nimrod Moiseyev$^{1,2}$, L.\ D.\ Carr$^{3,*}$, Boris A.
Malomed$^{4}$,
and Y. B. Band$^{5}$}
\address{
$^1$ Institute of Theoretical Atomic and Molecular Physics,
Harvard-Smithsonian Center for Astrophysics, Cambridge, MA 01238. \\
$^2$ Department of Chemistry and Minerva Center for Nonlinear Physics of
Complex Systems, Technion -- Israel Institute of Technology Haifa 32000,
Israel. \\
$^3$ Laboratoire Kastler Brossel, Departement de physique, Ecole normale
sup\'erieure, 24 rue Lhomond, 75231 Paris CEDEX 05, France \\
$^*$ Present address: JILA, National Institute of Standards and Technology and
Physics Department, University of Colorado, Boulder, 80309-0440 \\
$^4$ Department of Interdisciplinary Studies, Faculty of Engineering,
Tel Aviv University, Tel Aviv 69978, Israel \\
$^5$ Departments of Chemistry and Electro-Optics, Ben-Gurion University
of the Negev, Beer-Sheva, 84105 Israel}

\begin{abstract}
It is shown using the Gross-Pitaevskii equation that resonance states
of Bose-Einstein condensates with attractive interactions can be
stabilized into true bound states.  A semiclassical variational
approximation and an independent quantum variational numerical method
are used to calculate the energies (chemical potentials) and linewidths
of resonances of the time-independent Gross-Pitaevskii equation; both
methods produce similar results.  Borders between the regimes of
resonances, bound states, and, in two and three dimensions, collapse,
are identified.
\end{abstract}

\maketitle

Resonances are metastable quantum states which have finite lifetimes;
they play a key role in different types of scattering
phenomena~\cite{Taylor}.  Simon \textit{et al.}~\cite{Reed_Simon}
provided conditions for sharp transitions from a resonance to a true
bound state as a parameter in the system's Hamiltonian is varied.
These quantum phase transitions have been studied theoretically and
experimentally in the context of the linear Schr\"{o}dinger equation
\cite{sabree_review}.  Our objective here is to search for similar
transitions in nonlinear systems, such as Bose-Einstein condensates
(BECs)~\cite{Pitaevskii_review}.  In particular, we address the
question of whether resonance states of a BEC with attractive binary
atomic interactions can be stabilized into bound states by making the
effective nonlinearity increasingly negative (through increase of the
number of atoms, or tuning the $s$-wave scattering length $a_{s}$ near
a Feshbach resonance to be more negative).  As the existence and
stability of eigenfunctions of nonlinear wave equations is also a
central issue in other fields, and nonlinear Schr\"{o}dinger type
equations apply to many of them, including nonlinear optics, spin waves
in magnetic films, Langmuir waves in hot plasmas, and gravity surface
waves in fluids, the study of the effect of nonlinearity on resonances
is of general physical interest.

We develop two independent complementary approaches to this problem.
The first utilizes a variational Gaussian ansatz for the
quasi-stationary state and the WKB approximation to calculate the
tunneling rate.  The second utilizes a fully quantum variational method
with an absorbing potential, which, in the case of the linear
Schr\"odinger equation, may be shown formally \cite{riss1993} to be
equivalent to the exterior scaling transformation \cite{simon1979}.  We
note parenthetically that complex scaling has been previously used to
predict resonances for nonlinear Hartree-Fock models arising in
electronic structure calculations of atoms and molecules
\cite{Rescigno}.  In both approaches, we seek to determine the energies
and linewidths of resonance states of the Gross-Pitaevskii equation
(GPE), which is a nonlinear Schr\"{o}dinger equation (NLS), derived in
the mean-field approximation.  In the context of the BEC, the
\textquotedblleft energy\textquotedblright\ may be interpreted as the
chemical potential which is the real part of the eigenvalue associated
with a resonance state, \textit{i.e.}, $\mathrm{Re}(\mu) $.  The
\textquotedblleft linewidth\textquotedblright\ $\gamma $ may be
interpreted as the coefficient in the rate equation, $dN/dt=-\gamma
(N)N $, where $N$ is the number of trapped atoms in the BEC and $\gamma
(N)=-2\,\left( \mathrm{Im}(\mu) \right) /\hbar$ is the decay rate.  We
emphasize that $\mathrm{Im}(\mu) $ is a function of $N$: thus the
tunneling problem associated with the decay of the resonance is a
nonlinear one, and the lifetime of a resonance can be obtained by
integrating $dN/[-\gamma(N)N]$.  In the BEC, such nonlinear tunneling
problems have been studied theoretically for bright solitons
\cite{carr29} and experimentally in the context of the relaxation of
condensate spin domains \cite{stamperkurn}.

The two general methods are applied to the specific case of an external
potential in the form of a harmonic well multiplied by a Gaussian
envelope.  This choice of the potential is motivated by studies of
optically trapped BECs \cite{barrett2001}, where the waist of the
trapping laser beam creates a broad Gaussian envelope.  The central
harmonic well may be created by a second, more narrowly focused laser
beam, or by a higher Hermite-Gauss mode of the laser, as used in
creating BEC waveguides \cite{bongs2001}, or by a laser beam with
embedded vorticity induced by passing the beam through a phase mask
\cite{Heckenberg}.  The lifetime given by tunneling of a quasi-bound
resonant state can be reduced by other loss processes, such as
three-body recombination and interactions with the background gas in
the vacuum, as in the case for any trapped BEC. However, as these loss
processes are the same for both resonant and bound states, tunneling of
the mean field can cause a measurable reduction in lifetime.  An
external potential in the GPE is similar to the waveguiding profile
created by a transverse modulation of the refractive index that appears
in the generalized NLS used for modeling propagation of spatial beams
in optical media in effectively 1D (planar waveguide) and 2D (bulk)
geometries \cite{Agrawal}.  The potential that we introduce below
corresponds to the waveguiding profile that can be easily engineered in
optical media.

The time-independent isotropic GPE with such an external potential can
be written in terms of dimensionless variables scaled to the harmonic
oscillator frequency $\omega $ and length $\ell_{\mathrm{ho}} \equiv
\sqrt{\hbar /m\omega }$ as
\begin{eqnarray}
-\left[\Psi ^{\prime \prime }+{\left( D-1\right) }\Psi ^{\prime
}/r\right]/2 + V(r)\Psi +U_{0}|\Psi ^{2}|\Psi = \mu \Psi \,,
\label{eqn:GPE} \\
V(r) = \left( V_{0}+r^{2}/2\right) \exp \left( -\alpha r^{2}\right)\,,
\label{eqn:pot} \\
\beta _{D}\int_{0}^{\infty }dr\,r^{D-1}|\Psi |^{2} = 1 \,.
\label{eqn:norm}
\end{eqnarray}
Here $\Psi (r)$ is the normalized wave function, $\mu $ is the
above-mentioned complex eigenvalue, $\beta _{1,2,3}=2,2\pi ,4\pi $ for
the spatial dimension $D=1,2,3$, and $U_{0}\propto a_{s}N$ is the
appropriately scaled nonlinearity coefficient in dimension D
\cite{Tripp00}, with $a_{s}$ the $s$-wave scattering length.  The
prime stands for $d/dr$, and $\alpha \equiv (\ell
_{\mathrm{ho}}/\ell_{\mathrm{Gauss}})^{2}$ is a parameter which
characterizes the structure of the potential.  For a broad Gaussian
envelope, which pertains to the experimentally available optical trap
described above, $\alpha \ll 1$.

\textit{Variational WKB approach} --- Approximate solutions to
Eq.~(\ref{eqn:GPE}) with potential (\ref{eqn:pot}) can be obtained via
variational methods.  We adopt a Gaussian \textit{ansatz}, $\Psi
_{\mathrm{an}}(r) = A\,\exp [-r^{2}/(2\rho ^{2})]$, where the width
$\rho$ and amplitude $A$ are variational parameters.  Substituting the
ansatz into the normalization condition (\ref{eqn:norm}) and the
Lagrangian of the GPE (for the time being, $\mathrm{Im}(\mu)$ is
disregarded) and minimizing it with respect to $\rho$ and $A$, one
obtains $A^{2}=2/[\beta _{D}\rho ^{D}\Gamma (D/2)]$, and
\begin{eqnarray}
\mathrm{Re}(\mu)  &=&\left\{ (-4+D)/\rho ^{2}+(1+\alpha \rho
^{2})^{-2-D/2}\right.   \nonumber \\
&&\left.  \times \lbrack 4V_{0}+(4+D)\rho ^{2}-\alpha (D+4V_{0}\alpha
)\rho ^{4}]\right\} /4\,,\,\,\,\,\,\, \label{eqn:mu}
\end{eqnarray}
\begin{eqnarray}
|U_{0}| &=&\Gamma (D/2)2^{-2+D/2}\beta \rho ^{-2+D}(1+\alpha \rho
^{2})^{-2-D/2}  \nonumber \\
&\times& \left\{ -2(1+\alpha \rho ^{2})^{2+D/2}+\rho
^{4}[2-4V_{0}\alpha -\alpha (D+4V_{0}\alpha )\rho ^{2}]\right\} \,,
\label{eqn:U0}
\end{eqnarray}
where $\Gamma (D/2)$ is a Gamma function.  The solution is stable in
the framework of the time-dependent GPE if $d\mathrm{Re}(\mu
)/d|U_{0}|\leq 0$, which is known as the Vakhitov-Kolokolov
criterion~\cite{VK}.

One can now apply the WKB approximation to find $\mathrm{Im}(\mu)$
\cite{Landau}.  In the 1D case, it is given by the standard expressions
\begin{eqnarray}
\gamma  &=&\textstyle \exp ( -2\vert \int_{r_1}^{r_2}dr\,p(r)\vert
)/T \,, \\
p &=&\textstyle\sqrt{2[ \mu -V_{\mathrm{eff}}(r)]
}\,,~T=\textstyle4\,\vert
\int_{0}^{r_1}dr/p(r)\vert \,,
\end{eqnarray}
where $V_{\mathrm{eff}}(r) \equiv \lbrack V(r)+U_{0}|\Psi(r)|^{2}]$,
and the endpoints $r=r_1$ and $r=r_2$ are found from $p(r)=0$.  In the
2D and 3D cases, it is necessary to transform the GPE into a 1D form
via $\Psi = \phi /\sqrt{r}$ and $\Psi = \phi /r$, respectively (for the
spherical wave).  In the former case, this yields $V_{\mathrm{eff}}(r)
= V(r)-(1/8)r^{-2} + U_{0}r^{-1}|\phi ^{2}|$, whereas in the latter,
$V_{\mathrm{eff}}(r) = V(r) + U_{0}r^{-2}|\phi^{2}|$.  Note that
$\mathrm{Im}(\mu )=-\gamma /2$ in these units; also, $\gamma$ must be
multiplied by $2$ in the 1D case, to account for tunneling from both
sides of the well.  Results for the 1D, 2D, and 3D cases are
illustrated in the figures below.

\textit{Exterior Complex Scaling of the GPE} --- In the linear
Schr\"{o}dinger equation ($U_{0}=0$), resonances are associated with
complex eigenvalues of the Hamiltonian which are obtained by imposing
outgoing-wave boundary conditions on $\Psi$.  Such eigenvectors are
sometimes called Siegert states.  They are not in the Hilbert space,
hence the Hamiltonian is a \emph{non-Hermitian} operator over the class
of functions including the Siegert states.  In order to calculate
resonances, it is convenient to carry out a similarity transformation
which \textquotedblleft brings back\textquotedblright\ the resonances
to the Hilbert space.  A well-known transformation of this kind is
complex scaling, wherein the coordinates are rotated into the complex
plane by a fixed angle $\theta$ \cite{Nimrod_Review,Reinhardt_Review}.
When $\theta $ is sufficiently large, the resonance eigenfunctions
become square-integrable.  Although the complex eigenvalues are $\theta
$-independent, provided $\tan (2\theta )>\gamma {/} [2\mathrm{Re}(\mu
)]$, the corresponding resonance eigenfunctions do depend on $\theta $:
therefore the nonlinearity prevents direct application of
complex scaling to the GPE. However, with the use of exterior
\cite{simon1979} or smooth exterior scaling
\cite{NM_Hirschfelder,Volodya_NM}, the potential $V(r)$ remains
unscaled except at large $r$, and the small distortion of the effective
nonlinear potential, $U_{0}|\Psi |^{2}$, may be neglected.

It has been shown for the linear Schr\"odinger equation that that
smooth exterior scaling transformations are equivalent to adding a
reflection--free complex absorbing potential (CAP) to the
Hamiltonian~\cite{Volodya_NM}, which is often approximated by a local
negative imaginary potential.  We chose $V_{\mathrm{CAP}}(r)=-i\lambda
(r-r_{0})^{4}$, if $r>r_{0}$, and $V_{\mathrm{CAP}}(r)=0$, if $r\leq
r_{0}$, where $r_{0}$ is a point near the edge of the grid, and
$\lambda$ is a variational parameter (unrelated to the above
variational approximation).  This CAP is the most common one used in
calculations of resonances in atoms, molecules, and nuclei.  The value
of $\lambda$ is found numerically from the variational condition $d\mu
/d\lambda = 0$, where $\mu$ is generally complex, and stationary
solutions with respect to small variation of $\lambda$ are obtained;
note that stationary solutions in the $\lambda$-variational space
reguires optimization of another parameter, such as the box size in
physical space, $L$ (see the ``cusp'' condition in
\cite{Nimrod_Review}).  The value of $r_{0}$ is chosen so that the
wavefunction is exponentially small for $r>r_{0}$; this ensures that
the smooth exterior-scaling transformation does not distort the
effective potential $V_{\mathrm{eff}}(r)$ (for a detailed discussion of
this point, see \cite{riss1993,Volodya_NM} and references therein).

In our calculations, sinusoidal functions were used as a basis set
in the 1D and 3D cases, and Bessel functions in 2D. Starting from
zero, the nonlinear strength parameter $U_{0}$ was decreased in
steps of $\delta U_{0}<0$, so that the overlap between the
solutions for $U_{0}$ and $U_{0}+\delta U_{0}$ was close to unity,
to ensure that the evolution of a single resonance was followed.
Thus, resonances could be identified by the CAP method, at each
step of the iteration, in the same manner as in linear quantum
mechanics.

\begin{figure}[t]
\centerline{
\includegraphics[width=3in,angle=0,keepaspectratio]{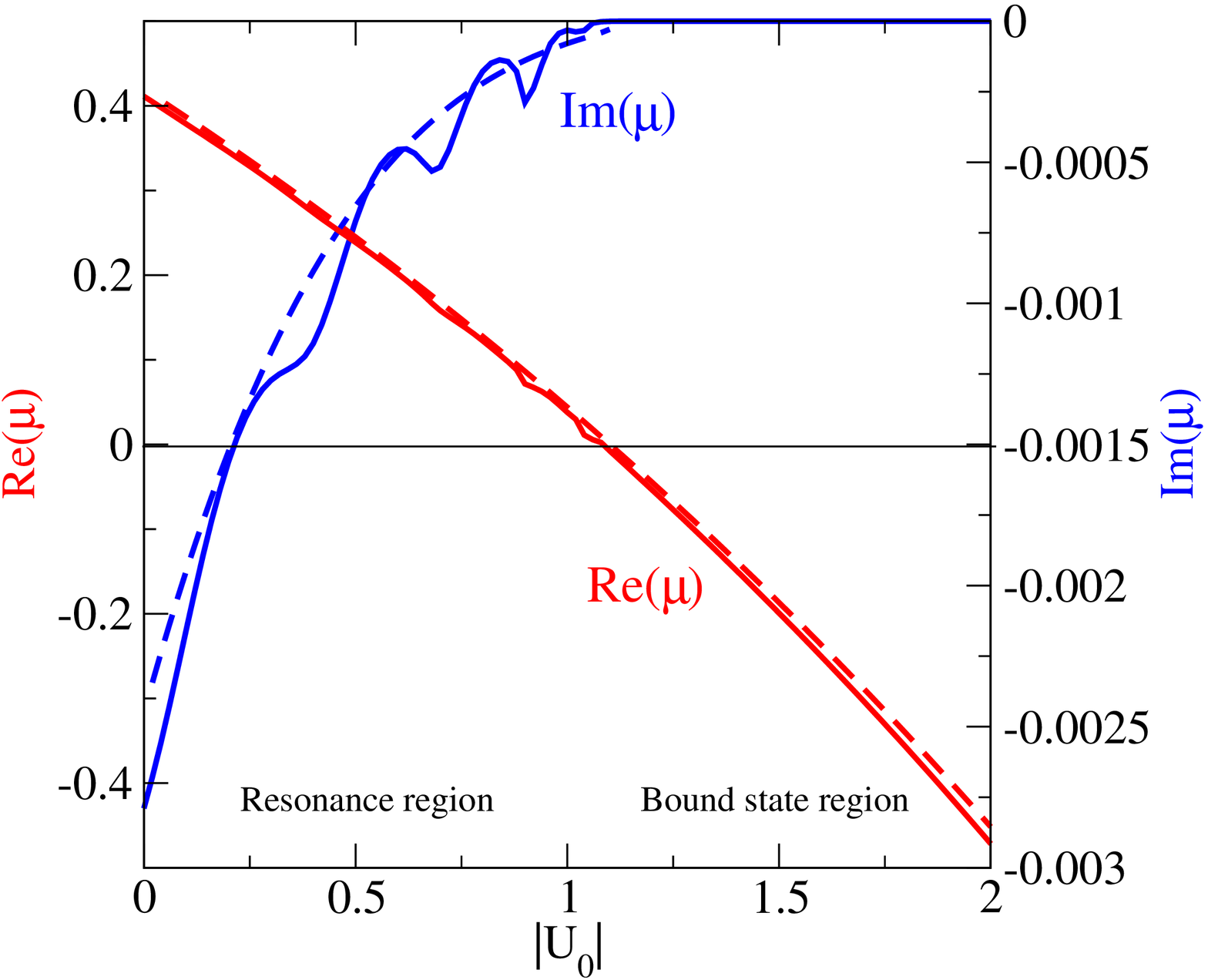}}
\caption{\textit{One dimension.} Shown is the
$\mathrm{Re}\protect(\mu)$ and $\mathrm{Im}\protect(\mu)$ versus
$|U_{0}|$ for the potential in the form of a harmonic well times
the Gaussian envelope [see Eq.~(\protect\ref{eqn:pot})], with
$\alpha\equiv (\ell_{ \mathrm{ho}}/\ell _{\mathrm{Gauss}})^{2} =
0.2$. Solid curves: results of the numerical method utilizing a
complex scaling method.  Dashed curves: the variational WKB
approximation. The critical point for the conversion of the
resonance into a bound state is $|U_{0}^{\mathrm{crit}}|=1.09$. }
\label{Fig1}
\end{figure}

\textit{GPE Resonances and Bound States} --- Consider first the case of
one dimension and no offset, {\it i.e.}, $D=1$ and $V_0=0$.  No bound
states and a single resonance exist in this potential for the
linear-Schr\"{o}dinger ($U_{0}=0$); the latter has $\mathrm{Re}(\mu) =
0.41134$ and $\mathrm{Im}(\mu) =-0.0027894$.  Beginning with this known
result, the $\mathrm{Re}(\mu)$ and $\mathrm{Im}(\mu)$ were calculated
numerically as a function of the nonlinear parameter $U_0$ by
adiabatically decreasing the nonlinear parameter from zero.  These
numerical results, and those obtained via the above analytical methods,
are displayed in Fig.~\ref{Fig1}.  The transition from the resonance to
a true bound state occurs at $U_{0}^{\mathrm{crit}} = -1.09$.  It is
noteworthy that, despite the well known fact that any 1D symmetric
potential with minimum below the asymptotic value of the potential
gives rise to at least one bound state in linear quantum mechanics, the
\emph{nonlinear} 1D GPE creates a bound state only if $-U_{0}\geq \vert
U_{0}^{\mathrm{crit}}\vert$ (i.e., if $U_{0}\leq
U_{0}^{\mathrm{crit}}$).

\begin{figure}[t]
\centerline{
\includegraphics[width=3in,angle=0,keepaspectratio]{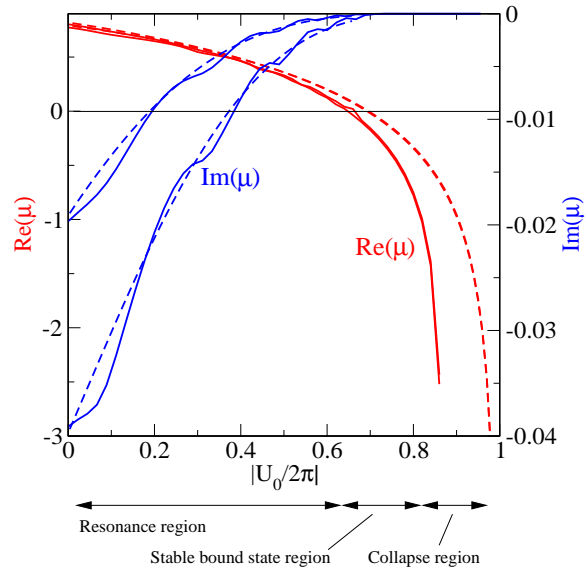}}
\caption{\textit{Two dimensions.}  Same as in
Fig.~\protect\ref{Fig1} for two different values of the
potential-shape parameter, $\protect\alpha = 0.16$ and
$\protect\alpha = 0.18$ (the upper and lower curves, respectively;
the analytical curves for $\mathrm{Re}\protect(\mu) $ at both
values of $\protect\alpha$ completely overlap).  Regions of the
resonance, bound state, and collapse are indicated.} \label{Fig2}
\end{figure}

In the 2D case with $V_{0}=0$, the numerical solution of
Eqs.~(\ref{eqn:mu}) and (\ref{eqn:U0}) shows that a resonance or
bound state can exist for $\alpha <\alpha _{c}=0.19245$, where the
tunneling rate diverges as $\alpha_{c}-\alpha \rightarrow +0$. For
$\alpha \geq \alpha_c$, the potential walls are too thin to retain
the condensate. The cases of $\alpha = 0.18$ and $\alpha = 0.16$
are shown in Fig.~\ref{Fig2}.  The numerical study demonstrates
that collapse occurs at the point $U_{0}^{\mathrm{coll}}/(2\pi
)=-0.86$, while the variational approximation gives a result
$13\%$ larger (because the above ansatz is a poor approximation
near the collapse, see Ref.~\cite{carr29} and references therein).
The transition from the resonance to a bound state takes place at
the point $U_{0}^{\mathrm{crit}}/(2\pi ) = -0.66$. These values
vary slightly with $\alpha$.

\begin{figure}[t]
\centerline{
\includegraphics[width=3in,angle=0,keepaspectratio]{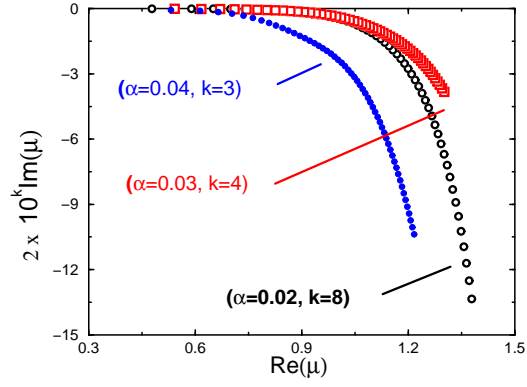}}
\caption{\textit{Three dimensions.} The width of the resonance states
versus energy for three different wells with $V(0)=V(\infty)$ (i.e.,
$V_{0}=0$): $\alpha \equiv (\ell_{\mathrm{ho}}
/\ell_{\mathrm{Gauss}})^{2} = 0.02$, $0.03$, and $0.04$.  For the
definition of $k$ see the label attached to the vertical axis.  In each
case, the collapse point is reached before the resonance can be
stabilized into a bound state.  The variational WKB approximation
produces similar results (not shown here).}
\label{Fig3}
\end{figure}

In the 3D case, we again start by considering $V_{0}=0$.  Varying
$\alpha$, we found that resonances never turn into bound states as the
nonlinearity coefficient $U_{0}$ is made more negative.  Instead, the
condensate collapses \cite{collapse}, as shown in Fig.~\ref{Fig3}; the
variational approximation produces exactly the same result (details
are not displayed here).  It is not surprising that collapse sets in
after the stabilization of the bound state in the 2D case
(Fig.~\ref{Fig2}), while this does not happen in 3D: as is well known,
the cubic nonlinearity gives rise to weak and strong collapse in 2D
and 3D, respectively (the nonlinear term of the effective potential
in 2D and 3D is $U_{0}r^{-1}|\phi|^{2}$ and $U_{0}r^{-2}|\phi|^{2}$,
respectively).

\begin{figure}[t]
\centerline{
\includegraphics[width=3in,angle=0,keepaspectratio]{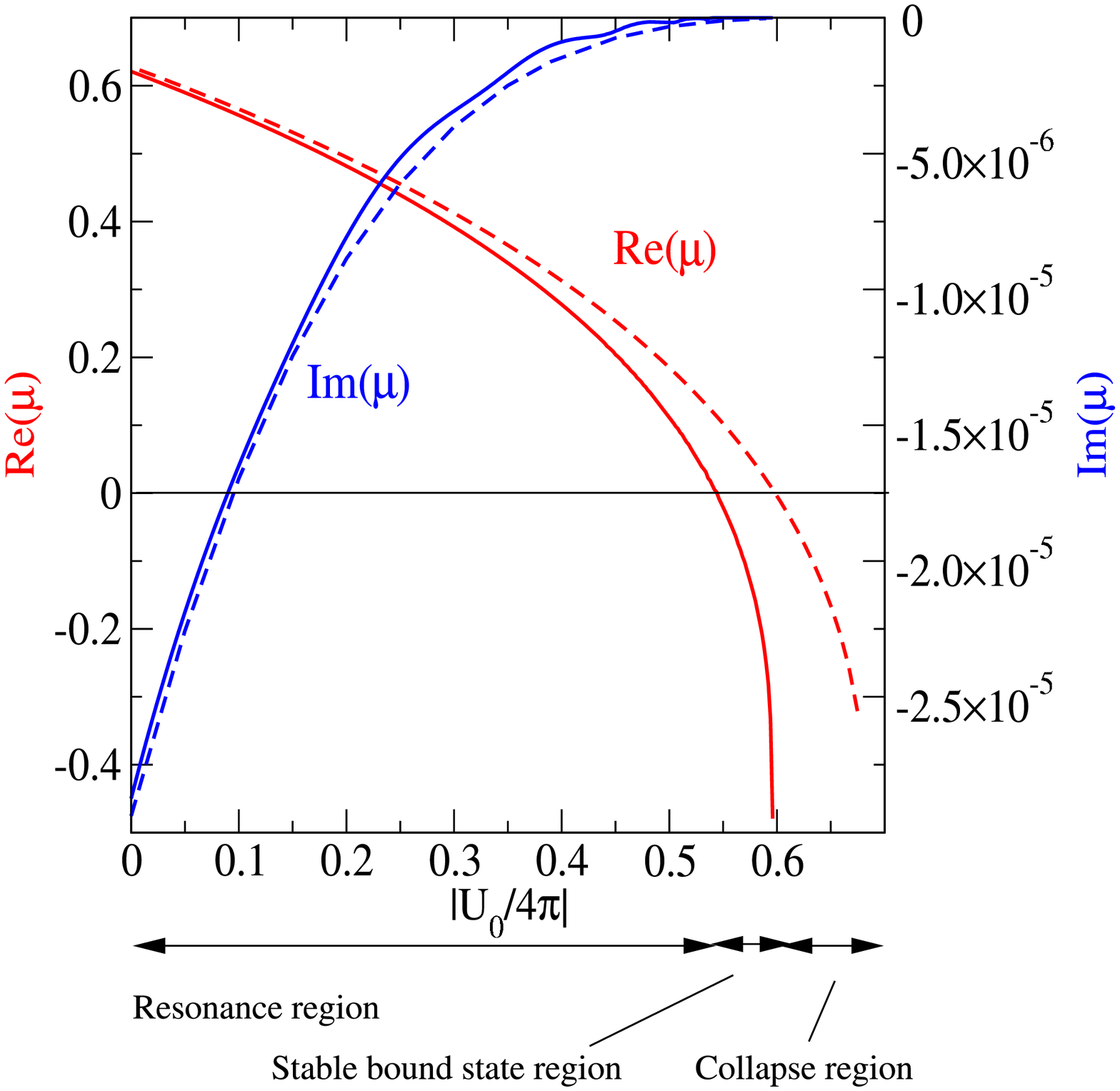}}
\caption{\textit{Three dimensions.}  Same as in Fig.~\protect\ref{Fig1}
for the potential (\protect\ref{eqn:pot}) with $\protect\alpha = 0.02$
and $V_{0} = -0.8$ (so that $V(0)=V(\infty )-0.8$).  The offset $V_{0}
< 0$ allows for the stabilization of the resonance into a bound state,
unlike the case shown in Fig.~\protect\ref{Fig3}.}
\label{Fig4}
\end{figure}

The above result for the 3D case may be understood as follows: the
variational approximation with the Gaussian ansatz shows that the
chemical potential at the collapse point in the isotropic 3D
harmonic well \textit{without} a Gaussian envelope is {$\mu
^{\mathrm{coll}}= 0.224$}.  Since this value is greater than zero,
and the full potential has $V(\infty) = 0$, it is not possible to
transform a resonance into a bound state.  However, lowering
$V(0)\equiv V_{0}$ [see Eq.~(\ref{eqn:pot})] with respect to
$V(\infty )$ makes such a transition possible.  To illustrate
this, the case of $V_{0}=-0.8$ and $\alpha =0.02$ is shown in
Fig.~\ref{Fig4}. The corresponding potential~(\ref{eqn:pot}) has a
single resonance and no bound states in the linear limit
($U_{0}=0$). The figure shows that, as $|U_{0}|$ increases, the
width and energy of the resonance decrease, and indeed the
resonance is stabilized into a bound state at the point
$U_{0}^{\mathrm{crit}}/(4\pi )=-0.552$.  As in the 1D and 2D
cases, the variational WKB approximation provides for reasonable
accuracy, as compared to the numerical results.

Clearly, our two independent methods may also be used to address the
question of whether BECs with \emph{repulsive} nonlinearity have a
critical point beyond which bound states trapped in a potential well
destabilize into resonances as the nonlinearity is increased.

In conclusion, we have demonstrated within a mean field approximation
that resonances in a BEC can be transformed into true bound states, as
the strength of the attractive nonlinearity increases beyond a critical
value $\vert U_{0}^ {\mathrm{crit}}\vert$.  Borders between three
different dynamical regimes, \textit{viz.}, resonances, bound states,
and collapse, have been delineated.  In 1D, resonances can always be
stabilized into bound states, as collapse is not possible.  In 2D, we
find that the transition of the resonance into a bound state can be
readily achieved before collapse occurs, as the collapse is weak in
this case.  However, in 3D, where collapse is strong, the critical
value for the transition, $|U_{0}^{\mathrm{crit}}|$, can be made
smaller than the collapse threshold, $|U_{0}^{\mathrm{coll}}|$, only if
$V(0)<V(\infty )$, \textit{i.e.}, the floor of the potential well is
lower than at infinity.  In optically trapped BECs, this can be
achieved by means of depression of the effective potential well with a
red-detuned laser.

\bigskip

\textit{Acknowledgments} --- We thank David Guery-Odelin and Jean
Dalibard for useful discussions.  This work was supported by NSF
through a grant at ITAMP, NSF through grant No.~MPS-DRF 0104447,
Israel Science Foundation (grant No.~212/01), Israel Ministry of
Defense Research and Technology Unit, Israel Science Foundation for a
Center of Excellence (grant No.~8006/03) and by the Binational
(US-Israel) Science Foundation (grant Nos.~1999459 and 2002147).

\newpage

\begin{thebibliography}{99}
\bibitem{collapse} What actually happens in the case of collapse cannot
be determined within the mean-field description.  We do not address
this issue here.
\end{thebibliography}
\end{document}